\documentclass[aps,preprint,nofootinbib,eqsecnum]{revtex4}
\usepackage{amsmath,amssymb,bm}
\usepackage{graphicx}
\usepackage{epstopdf}
\usepackage{latexsym}

\begin{document}
\preprint{arXiv:0808.0191}

\title{Quantum phase transitions beyond the Landau-Ginzburg paradigm and supersymmetry}

\author{Subir Sachdev and Xi Yin}
\affiliation{Department of Physics, Harvard University, Cambridge MA
02138}
\date{August 1, 2008 \\
\vspace{1.6in}}

\begin{abstract}
We make connections between studies in the condensed matter literature
on quantum phase transitions in square lattice antiferromagnets, and results in the particle
theory literature on abelian
supersymmetric gauge theories in 2+1 dimensions. In particular, we point out
that supersymmetric $U(1)$ gauge theories (with particle content similar, but not identical, to those
of theories of doped antiferromagnets) provide rigorous examples of 
quantum phase transitions which do not obey the Landau-Ginzburg-Wilson
paradigm (often referred to as transitions realizing 
``deconfined criticality''). We also make connections between supersymmetric mirror symmetries
and condensed matter particle-vortex dualities.
~\\
\end{abstract}

\maketitle

\section{Introduction}
\label{sec:intro}

The condensed matter literature has seen much discussion \cite{senthil1,senthil2,mv,sandvik,melkokaul,wiese,kuklov,mv2,flavio1} on quantum
phase transitions that violate the Landau-Ginzburg-Wilson (LGW) paradigm. In such transitions,
conventional phases with distinct broken symmetries are generically separated by a
second-order quantum critical point {\em e.g.\/} a theory with a global symmetry group
$G_1 \times G_2$ has a transition from a phase where $G_1$ is broken while $G_2$ is
preserved, to a phase where $G_1$ is preserved and $G_2$ is broken. The theory
for the critical point is usually not expressed in terms of order parameters which measure
$G_1$ or $G_2$ symmetry breaking, but in terms of `fractionalized' degrees of freedom,
and hence the terminology `deconfined'.

Non-LGW transitions have been known in 1+1 dimensions for some time \cite{oned}, and our interest
in this paper is exclusively in 2+1 dimensions. Examples have been established \cite{rs1,rs2,rs3,senthil1,senthil2} for $SU(N_f)$
square lattice antiferromagnets for large, but finite, $N_f$, as we shall review below. The situation for the physically
most interesting case of $N_f=2$ remains unsettled: while there is convincing evidence for the emergent degrees of freedom
of the deconfined field theory, there are open questions on the nature of the critical point, with some results favoring a first-order transition. \cite{sandvik,melkokaul,wiese,kuklov,flavio1,mv2}

Although this does not appear to have been recognized in the condensed matter literature,
examples of non-LGW transitions have also appeared in the particle theory literature:
they are present in supersymmetric field theories in 2+1 dimensions
studied in the
early work of Seiberg and collaborators \cite{sw,is,ks}. The purpose of our paper is to review
the condensed matter and particle theory results in a unified manner. Our aim is to make our
discussion intelligible across the boundaries of these fields. We hope to convince the reader
that there is a remarkably close analogy between supersymmetric `deconfined criticality' and the models
arising in the study of quantum antiferromagnets. A close connection will also be drawn between
the `mirror symmetry' of the supersymmetric field theories and particle-vortex duality arguments.

An important field theory arising \cite{rs3,senthil1,senthil2} in the study of the loss of N\'eel order in $SU(N_f)$ antiferromagnets
on the square lattice is the Abelian Higgs model with $N_f$ complex scalar fields, $q_i$, $i=1 \ldots N_f$,
with the Euclidean Lagrangian
\begin{equation}
\mathcal{L}_H (q_i, A_\mu)  = |(\partial_\mu - i A_\mu )q_i|^2 + r |q_i|^2 + u (|q_i|^2)^2 + \frac{1}{g^2} F_{\mu\nu}^2
\label{fh}
\end{equation}
where $A_\mu$ is a $U(1)$ gauge field, and $F_{\mu\nu} = \partial_\mu A_\nu - \partial_\nu A_\mu$
is the `electromagnetic' field. Here, and henceforth, all Lagrangians are assumed to be integrated
over 2+1 dimensional spacetime to obtain the associated actions. For some purposes, it is useful
consider a `strong-coupling' limit of $\mathcal{L}_H$ in which the quartic potential is replaced by a hard constraint
$\sum_i |q_i|^2 = 1$: in this case $\mathcal{L}_H$ describes the so-called NCCP$^{N_f - 1}$ model \cite{mv2}
(for non-compact $U(1)$ gauge field CP$^{N_f - 1}$ model). The universal properties of these two cases are expected to be identical.

For $r$ sufficiently negative, $\mathcal{L}_H$ is in the Higgs phase, and the global $SU(N_f)$ symmetry
is broken because of the condensation of the $q_i$ (for $N_f > 1$).
The gauge-invariant `meson' operators $q_i^\ast q_j$ are the order parameters for this symmetry
breaking: these constitute the N\'eel order parameters of the $SU(N_f)$ antiferromagnet.

For $r$ sufficiently positive, $SU(N_f)$ symmetry is restored, and $q_i$ are massive scalar particles
which interact by exchanging $A_\mu$ photons. The electric potential between these scalars has
the `Coulomb' form of $\ln (r)$, and so this is referred to as the Coulomb phase.

There is a great deal of interest on the nature of the transition between the Higgs and Coulomb phases as a function
of increasing $r$. For $N_f=1$ the transition can be either first or second order depending upon
parameters; for large $N_f$ it is second order; and for $N_f=2$ the question is the focus of the recent debate.\cite{sandvik,melkokaul,wiese,kuklov,flavio1,mv2}

Crucial to our purposes here is the fact that $\mathcal{L}_H$ enjoys an additional global symmetry, distinct
from the $SU(N_f)$ flavor symmetry. This global symmetry is special to 2+1 dimensions, and is linked to the presence
of the `topological' current
\begin{equation}
\widetilde{J}_\mu = \frac{1}{4\pi} \epsilon_{\mu\nu\lambda} F_{\nu \lambda}
\end{equation}
which is conserved, $\partial_\mu \widetilde{J}_\mu = 0$, reflecting the conservation of total magnetic flux.
We can associate this conservation law with a dual $\widetilde{U}(1)$ global symmetry, and consequently a field operator which changes the magnetic flux will carry $\widetilde{U}(1)$ charge. Because of the presence of the fields $q_i$ with unit electrical charge,
the Dirac quantization condition implies that the magnetic flux can only change in integer multiples of $2 \pi$. We can therefore
introduce the elementary monopole creation operator $\hat{q}$ which carries unit $\widetilde{U}(1)$ charge; these
are `topological disorder operators' for the $U(1)$ gauge theory \cite{murthy,kapustin,kovner,max}. As written, the continuum theory $\mathcal{L}_H$ does not allow for the creation
of any monopoles, and this is equivalent to the existence of the
global $\widetilde{U}(1)$ symmetry. We can now ask for the fate of this $\widetilde{U}(1)$ symmetry by examining
the two-point monopole correlator in both phases. It is not difficult to show that as $|{\bf r}| \rightarrow \infty$
\begin{equation}
\langle \hat{q} ({\bf r}) \hat{q}^\dagger (0) \rangle \sim \left\{
\begin{array}{ccc}
\exp( - m |{\bf r}| ) &,& \mbox{Higgs phase} \\
\mbox{const} &,& \mbox{Coulomb phase}
\end{array} \right.
\end{equation}
where $m$ is an energy scale characterizing the Higgs phase, of order the Higgs mass. From this it is clear that
there are long-range correlations in monopole operator in the Coulomb phase, and consequently the $\widetilde{U}(1)$ symmetry
is broken.

To summarize, the Abelian Higgs model has a global $SU(N_f)$ $\times \widetilde{U}(1)$ symmetry.
In the Higgs phase, the $SU(N_f)$ symmetry is broken and the $\widetilde{U}(1)$ symmetry is preserved,
while in the Coulomb phase  the $SU(N_f)$ symmetry is preserved and the $\widetilde{U}(1)$ symmetry is broken.

These conclusions appear to satisfy the requirements of a non-LGW transition, as defined above. However, an objection
might be raised that the $\widetilde{U}(1)$ symmetry is `topological', involves highly non-local transformations
of the field operators in $\mathcal{L}_H$, and so is not directly observable. A remarkable fact of the mapping between $\mathcal{L}_H$ and the $SU(N_f)$ quantum antiferromagnets is that the non-local monopole operator can be related
to simple local observables expressed in terms of the underlying lattice $SU(N_f)$ spins \cite{rs3,senthil2}.
In particular, for a class of $SU(N_f)$
antiferromagnets (with fundamental matter on the square lattice sites),
$\hat{q}$ is proportional to the `valence bond solid' (VBS) operator.
Further, the `hidden' $\widetilde{U}(1)$ symmetry is not hidden at all, but an enlargement of the spatial $Z_4$ rotation
symmetry of the square lattice.
These connections have been reviewed
in other recent articles \cite{ssnature,ssberlin}, and so we will not describe them further here. All we need for our purposes
is the conclusion that the $\widetilde{U}(1)$ symmetry is physical and experimentally measureable, and so the
Abelian Higgs model does indeed satisfy the conditions for a non-LGW transition.

For completeness, another significant feature of the connection between $\mathcal{L}_H$ and quantum antiferromagnets
should be mentioned. While the theory $\mathcal{L}_H$ does not permit any monopoles in the strict continuum limit,
the fate of the monopoles can only be correctly addressed by the actual short distance physics, which is that of the lattice antiferromagnet.
Here it is found that Berry phases \cite{haldane} lead to large short-distance cancellations between monopoles, and so an additive
contribution to the Lagrangian,  $\mathcal{L}_m \sim \hat{q} + \hat{q}^\dagger $
does not appear in the continuum theory \cite{rs3}. By a careful symmetry analysis of the Berry phases, it was shown \cite{rs3}
that the simplest allowed monopole term was  $\mathcal{L}_m \sim \hat{q}^4 + \hat{q}^{\dagger4} $. Recalling that
the $\hat{q}$ carries $\widetilde{U}(1)$ charge, we see that the actual magnetic symmetry
of the full theory $\mathcal{L}_H + \mathcal{L}_m$ is not $\widetilde{U}(1)$, but $Z_4$ (with is identified
with the square lattice rotation symmetry).
It is this $Z_4$ symmetry which is broken in the `Coulomb' phase.
However, it has been argued that such a $\hat{q}^4$ term is likely
irrelevant near the critical point \cite{senthil1,senthil2},
and this is supported by numerical studies \cite{sandvik,melkokaul,ssnature}. So $\mathcal{L}_m$ can be neglected in
the immediate vicinity of the critical point, and we will not consider it further here.

The remainder of this paper will describe the analogy between the properties of the above Abelian Higgs model with
$N_f$ scalars, and the corresponding model with $\mathcal{N}=4$ supersymmetry. Briefly, we simply promote the
fields of $\mathcal{L}_H$ to the corresponding $\mathcal{N}=4$ multiplets: we promote the scalars $q_i$ to
hypermultiplets $\mathcal{Q}_i$, and the $U(1)$ gauge field $A_\mu$ to a $U(1)$ vector multiplet $\mathcal{V}$. With $\mathcal{N}=4$
supersymmetry, the resulting theory is unique and has only a single dimensionful gauge coupling constant $g$;
a complete and explicit form of the Lagrangian appears in Section~\ref{sec:lag}.
We will explore its phase diagram ({\em i.e.\/} moduli space) and find remarkable analogies with the
non-supersymmetric $\mathcal{L}_H$.

This generalization to supersymmetric models necessarily involves introduction of Dirac fermions, both in the matter
and gauge multiplets. While precisely this field content is not known to be present in any models of interest in condensed matter,
recent work \cite{acl,flavio2,rkkss} has considered a theory in which Dirac fermions, $\psi_j$ ($j=1\ldots N_d=4$) with a $U(1)$ charge are added to $\mathcal{L}_H$ to obtain
\begin{equation}
\mathcal{L}_{H\psi} (q_i, \psi_j, A_\mu)  = |(\partial_\mu - i A_\mu )q_i|^2 + r |q_i|^2 + u (|q_i|^2)^2 + \frac{1}{g^2} F_{\mu\nu}^2 + \bar{\psi}_j \sigma^\mu ( \partial_\mu - i A_\mu ) \psi_j,
\label{fhpsi}
\end{equation}
where $\sigma^\mu$ are the Pauli matrices; these Dirac fermions represent the Bogoliubov quasiparticle excitations of a $d$-wave superconductor.
Note the number of scalars ($N_f$) and fermions ($N_d$) are not equal in the physical case, although this will be the
case in the supersymmetric models below.
The model $\mathcal{L}_{H\psi}$ is also expected to exhibit a non-LGW transition, which will then be even closer to the supersymmetric models.

We will begin in Section~\ref{sec:ah} by a review of the `particle-vortex' or `Dasgupta-Halperin' duality \cite{peskin,dasgupta}
of the Abelian Higgs model
with $N_f =1 $. Then, in Section~\ref{sec:ks} we will consider the corresponding model with $\mathcal{N}=4$ supersymmetry and $N_f=1$. We will show that the formulation of mirror symmetry for this model by Kapustin and Strassler \cite{ks} is precisely the same
as the exact statement of Dasgupta-Halperin duality in Section~\ref{sec:ah}. The analogy between these models also extends to the $N_f>1$
case, and this is discussed in Section~\ref{sec:multiple}.
The subsequent sections will explore the structure of the phases of the $\mathcal{N}=4$ theory.

\section{Abelian Higgs model for $N_f = 1$}
\label{sec:ah}

Here we will present the exact statement of duality properties of the Abelian Higgs model. We consider
only the case $N_f=1$, with comments about $N_f>1$ in Section~\ref{sec:multiple} below.

We write the generating function for the fluxes of the Abelian Higgs model as
\begin{equation}
Z_H [\hat{A}_\mu] = \int \mathcal{D} q \mathcal{D} A_\mu \exp \left( - \int  \left[
\mathcal{L}_H (q, A_\mu)  + \frac{1}{2 \pi} \epsilon_{\mu\nu\lambda} \hat{A}_\mu \partial_\nu A_\lambda  \right] \right).
\label{zh}
\end{equation}
In general, the functional $Z_H [\hat{A}_\mu]$ will depend upon all the couplings $r$, $u$, and $g$, and also on the
ultraviolet cutoff. However, in the vicinity of a second-order Higgs-Coulomb transition, the functional is dominated
by momenta much smaller than the cutoff, and depends only upon universal correlations of the conformal field
theory (CFT) describing this transition. The renormalization group fixed point describing the CFT is expected
to have only one relevant perturbation, whose strength is characterized by a single mass scale $m$ (which we have
to define separately in the Higgs and Coulomb phases). In the scaling limit of the fixed point, the functional
$Z_H [\hat{A}_\mu]$ is a {\em universal\/} dimensionless functional of $\hat{A}_\mu$, its momenta $p_\mu$, and the mass $m$,
all of which have unit scaling dimension. There are no known exact results for this functional, but approximate results
can be obtained using the $(4-D)$ expansion ($D$ is the dimensionality of spacetime), the $1/N_f$ expansion,
and by numerical simulations.

In the particle-vortex duality, the flux lines of $\mathcal{L}_H$ are mapped onto the world lines of dual particles
(`vortices') which are created by the monopole operator $\hat{q}$. Dasgupta and Halperin argued that these
dual particles only have short-range interactions, and so are described by the XY model -- in the continuum
limit, this is the theory $\mathcal{L}_H$, but with the no $A_\mu$ fluctuations.
The generating functional of the dual particle currents in such a theory is given by
\begin{equation}
Z_{XY} [ \hat{A}_\mu ] =  \int \mathcal{D} \hat{q}  \exp \left( - \int  \mathcal{L}_H ( \hat{q}, \hat{A}_\mu ) \right) \label{xy}
\end{equation}
Now we implicitly assume that the action $\mathcal{L}_H$ has different couplings $\hat{r}$, $\hat{u}$, and
$\hat{g}$. The XY model is known to have a second-order critical point, and just as for Eq.~(\ref{zh}),
the functional $Z_{XY} [\hat{A}_\mu]$ becomes completely universal in the scaling limit near the fixed point,
and depends only upon a single mass scale $\hat{m}$ measuring the deviation from the critical point.

We will now connect the generating functionals in Eqs.~(\ref{zh}) and (\ref{xy}). In general, this requires us to map
the couplings $r$, $u$ and $g$ to $\hat{r}$, $\hat{u}$, and $\hat{g}$. The form of this mapping is not
known, and indeed, depends upon the specific ultraviolet cutoff. However, the central hypothesis of Dasgupta and Halperin \cite{dasgupta}
was that the RG fixed points describing the transitions in the $N_f=1$ Abelian Higgs model and the XY model
are the same. Thus, in the scaling limit near this fixed point, to map the two theories to each other, we need only connect
the mass scales $m$ and $\hat{m}$. The latter is easily done, by identifying these with the corresponding particles/vortices
in the spectra. After this identification, we then have the remarkable exact duality statement
\begin{equation}
Z_H [ \hat{A}_\mu] = Z_{XY} [ \hat{A}_\mu ]. \label{dh}
\end{equation}
This is one of the very few exact statements about non-supersymmetric CFTs in 2+1 dimensions; it has not been rigorously
established, but the arguments based upon dualities of lattice models appear quite robust.

\section{Abelian gauge theory for $N_f=1$ with $\mathcal{N}=4$ supersymmetry}
\label{sec:ks}

The statements in Section~\ref{sec:ah} have a remarkably precise analog in $U(1)$ gauge theories
with $\mathcal{N} = 4$ supersymmetry, as stated by Kapustin and Strassler \cite{ks} (hereafter referred
to as KS).

As we discussed in Section~\ref{sec:intro}, we generalize the scalar $q$ to a $\mathcal{N}=4$
hypermultiplet $\mathcal{Q}$, and the gauge field $A_\mu$ to a $\mathcal{N}=4$ vector
multiplet $\mathcal{V}$. The precise field content of these hypermultiplets will be specified
later in Section~\ref{sec:lag}.

Now we consider the $\mathcal{N}=4$ theory with one hypermultiplet $\mathcal{Q}$
and one vector multiplet $\mathcal{V}$. This theory has a unique Lagrangian, $\mathcal{L}_S$, with only
one dimensionful gauge coupling constant $g$. This Lagrangian is the analog of Eq.~(\ref{fh})
and its explicit form of this will appear in Section~\ref{sec:lag}.
Next we define a generating functional for the `fluxes' of $\mathcal{V}$ which is the
analog of Eq.~(\ref{zh}) (as in Eq. (11) of KS):
\begin{equation}
Z_{\rm{SQED-1}} [\hat{ \mathcal{V}} ] = \int \mathcal{D} \mathcal{V} \mathcal{D} \mathcal{Q} \exp \left(-
\int \left[ \mathcal{L}_S ( \mathcal{Q}, \mathcal{V}) + \mathcal{L}_{BF} ( \hat{\mathcal{V}}, \mathcal{V} ) \right] \right)
\end{equation}
where $\mathcal{L}_{BF}$ is the analog of the Chern-Simons coupling in Eq.~(\ref{zh}) between
two vector fields (see Eq. (8) of KS), and we have dropped the gauge-fixing term included by KS.
An advantage of the supersymmetric theory is that it is easy to take the scaling limit associated
with the CFT: we simply send the gauge coupling $g \rightarrow \infty$. The subtle RG renormalizations of couplings
required for the non-supersymmetric case are not required here. Indeed, we may view this ultraviolet insensitivity
as the main crutch that is provided by supersymmetry; the infrared physics otherwise remains similar
to the non-supersymmetric case.

Finally, the statement of duality is just as in Eqs.~(\ref{xy}) and (\ref{dh}). We define (as in Eq. (12) of KS) the generating
function of currents of a dual monopole field $\mathcal{Q}$ by
\begin{equation}
Z_{\hat{\mathcal{Q}}} [ \hat{\mathcal{V}}] = \int \mathcal{D} \hat{\mathcal{Q}} \exp \left( - \int \mathcal{L}_H ( \hat{\mathcal{Q}}, \hat{\mathcal{V}}) \right)
\end{equation}
and then we have one of the main results of KS
\begin{equation}
Z_{\rm{SQED-1}} [\hat{ \mathcal{V}} ] = Z_{\hat{\mathcal{Q}}} [ \hat{\mathcal{V}}]. \label{zzss}
\end{equation}
Here, the universal scaling limits of the two sides are taken simply by the limit $g \rightarrow \infty$.
Unlike the non-supersymmetric case, the integrals over the hypermultiplets in $Z_{\rm{SQED-1}} [\hat{ \mathcal{V}} ]$  and $Z_{\hat{\mathcal{Q}}} [ \hat{\mathcal{V}}]$ are Gaussian, and so can be expressed as superdeterminants---this leads
to the key identity in Eq. (13) of KS. We will present a detailed illustration of the result (\ref{zzss}) in Section~\ref{sec:nf1}.

\section{$N_f > 1$}
\label{sec:multiple}

Now we briefly introduce the dualities for $N_f > 1$. Further discussion of the supersymmetric dualities appears
in the sections below.

For the non-supersymmetric case, statements of
dualities are only known for an `easy-plane' extension of $\mathcal{L}_H$ \cite{mv,bbbss}. In this case, we add an
additional quartic potential to the Lagrangian, {\em e.g.\/} $\sum_{i} |q_i|^4$, so that the continuous global
symmetry $SU(N_f)$ symmetry is reduced to $U(1)^{N_f -1}$, along with additional discrete symmetries.
The dual theory has $N_f$ scalar fields $\hat{q}_i$ and $N_f-1$ $U(1)$ gauge fields, such that $\hat{q}_i$
has charge $+1$ under the $i^{\rm th}$ $U(1)$ gauge field and charge $-1$ under the $(i-1)^{\rm th}$ $U(1)$ gauge field
(this is a quiver gauge theory).
The identity Eq.~(\ref{dh}) has a straightforward generalization to this case.

Note that the particle content of the dual
theory is identical to that of the direct theory only for $N_f=2$, and it was therefore conjectured that the CFT of the Higgs-Coulomb
transition is self-dual \cite{mv,m2cft}.

These dualities generalize to $\mathcal{N}=4$ supersymmetry, as reviewed by KS. However, now the dualities
apply also when there is full $SU(N_f)$ flavor symmetry.
A theory of $N_f$ hypermultiplets $\mathcal{Q}_i$ coupled to a vector multiplet $\mathcal{V}$ (this is the theory SQED-$N_f$)
is dual to a theory of $N_f$ hypermultiplets $\hat{\mathcal{Q}}_i$ coupled to $N_f-1$ vector multiplets, as proven
in Section III.C of KS.
These results for $\mathcal{N}=4$ supersymmetry were initially obtained by Intrilligator and Seiberg \cite{is},
and are described in their Sections 3.1 and 3.2 for $N_f = 2$ and $N_f > 2$ respectively.
In addition to the global $SU(N_f)$ flavor symmetry, these theories have a certain $SU(2)_L \times SU(2)_R$ symmetry required by $\mathcal{N}=4$
supersymmetry. Moreover, there is a global
$\widetilde{U}$(1) symmetry associated with the $U(1)$ gauge invariance in the direct formulation, just as discussed for
$\mathcal{L}_H$ in Section~\ref{sec:intro}. So the full symmetry is $SU(N_f) \times \widetilde{U}(1) \times SU(2)_L \times SU(2)_R$. We will see below
that the $SU(N_f)$ symmetry is broken (preserved) and the $\widetilde{U} (1)$ symmetry
is preserved (broken) in the Higgs (Coulomb) phase, as required for a non-LGW transition.

Note again that the particle contents of the direct and dual theories are identical for $N_f=2$. In this case the self-duality
of the CFT was established by Intrilligator and Seiberg \cite{is}. They also showed that the global symmetry of the CFT is enhanced
to $SU(2_f) \times \widetilde{SU}(2) \times SU(2)_L \times SU(2)_R$, with the first two $SU(2)$ factors exchanging under duality.

\section{Lagrangian with $\mathcal{N}=4$ supersymmetry}
\label{sec:lag}

This section will review the matter content of the $\mathcal{N}=4$ multiplets,
and the full form of the Lagrangian $\mathcal{L}_S$.

First, let us reduce to $\mathcal{N}=2$ superfields.
Each $\mathcal{N}=4$
hypermultiplet $\mathcal{Q}_i$ consists of a pair of $\mathcal{N}=2$ chiral superfields
$\mathcal{Q}_i = (Q_i, \tilde{Q}^i)$.
Each $\mathcal{N}=4$ vector multiplet, $\mathcal{V}$, consists of a $\mathcal{N}=2$ vector
superfield $V$ and a $\mathcal{N}=2$ chiral superfield $\Phi$.
The Euclidean
Lagrangian in ${\cal N}=2$ superspace (without the so-called Fayet-Illiopoulos (FI) term or masses) is
\begin{equation}
\mathcal{L}_S (\mathcal{Q}_i, \mathcal{V})  = \int d^4\theta ({1\over 4 g^2}\Sigma^2+ \frac{1}{g^2} \bar\Phi \Phi + \bar Q_i
e^{V} Q_i
+\bar {\tilde Q}^i e^{-V} \tilde Q^i) + \int d^2\theta \Phi Q_i\tilde Q^i + c.c.
\end{equation}
where $i=1 \ldots N_f$,  $\theta$ are superspace co-ordinates,
$\Sigma = \epsilon^{\alpha\beta} D_\alpha \bar D_\beta V$, and $D_\alpha$ are superderivatives.
See Appendix A of Ref.~\onlinecite{kleb} for a review of superspace notation in $D=3$.

Turning finally to the explicit components,
$V$ consists of
$A_\mu$, a scalar $\sigma$, the auxiliary field $D$ (which is integrated out),
and gaugino $\lambda$.
$\Phi$ consists of a complex scalar $\phi$ and its fermionic
partner $\psi$, as well as the auxiliary field $F$.
$(Q_i,\tilde Q^i)$ have components $(q_i, \tilde q^i)$ complex scalars,
$(\psi_{q_i},\psi_{\tilde q^i})$ Dirac fermions, and auxiliary fields
$(F_{q_i}, F_{\tilde q^i})$.
The full spacetime Lagrangian is then
\begin{eqnarray}
\mathcal{L}_S &=& \frac{1}{g^2} \left[ {1\over 4}F_{\mu\nu}^2 + {1\over 2}(\partial_\mu
\sigma)^2 +{1\over 2}D^2
+ \bar\lambda \sigma^\mu \partial_\mu
\lambda + |\partial_\mu \phi|^2 +|F|^2 + \bar\psi \sigma^\mu \partial_\mu\psi \right]
\nonumber \\
&&+|\nabla_\mu q_i|^2 + |\nabla_\mu \tilde q^i|^2
+ |\sigma q_i|^2 + |\sigma \tilde q^i|^2 + D(|q_i|^2-|\tilde q_i|^2)
+ |F_{q_i}|^2 + |F_{\tilde q^i}|^2 \nonumber \\
&&+ \bar\psi_{q_i} \sigma^\mu \nabla_\mu \psi_{q_i}+ \bar\psi_{\tilde q^i}
\sigma^\mu \nabla_\mu \psi_{\tilde q^i}+
\bar\psi_{q_i} \sigma \psi_{q_i}- \bar\psi_{\tilde q^i} \sigma \psi_{\tilde
q^i}\nonumber \\ & &+ i\bar q_i \bar\lambda \psi_{\tilde q^i}
-i \bar \psi_{\tilde q^i} \lambda q_i -i\bar {\tilde q}_i \bar\lambda \psi_{q_i}
+i \bar \psi_{q_i} \lambda \tilde q^i \nonumber \\
&&+ \left( F q_i \tilde q^i + \phi F_{q_i} \tilde q^i + \phi q_i F_{\tilde q^i}
+ \phi \psi_{q_i} \psi_{\tilde q^i} + \psi q_i \psi_{\tilde q^i} + \psi
\psi_{q_i} \tilde q^i +c.c. \right)
\end{eqnarray}
where $\nabla_\mu \equiv \partial_\mu - i A_\mu$.
Integrating out $D$, $F$, and $F_{q_i}, F_{\tilde q^i}$ yields the potential
terms
\begin{equation}
 V = {1\over 2}(\sum_i |q_i|^2 - \sum_i|\tilde q^i|^2)^2 + |\sum_i
q_i\tilde q_i|^2
+ \sum_i (|\phi q_i|^2 + |\phi \tilde q^i|^2) \label{effpot}
\end{equation}
Let us write
\begin{equation}
 q_{iA} = ( q_i, -\bar{\tilde q^i}),~~~~\psi_{ia} =
(\psi_{q_i},-\overline \psi_{\tilde q^i})
\end{equation}
where $a,A=1,2$ are $SU(2)_L\times SU(2)_R$ spinor indices.

We will denote the $SU(2)_L$ triplet $(\sigma,\phi=\phi_1+i\phi_2)$ by
the real symmetric field $\phi^{(ab)}$. The gauginos
$(\lambda,\psi)$ transform as $({\bf 2},{\bf 2})$ under $SU(2)_L\times SU(2)_R$,
and will be denoted by
$\chi_{aA}$. They satisfy the reality condition $(\chi_{aA})^\dagger=
\chi^{aA}=\epsilon^{ab}\epsilon^{AB} \chi_{bB}$.

Now the total scalar potential term can be written as
\begin{equation}
 V={1\over 2}\sum_{I=1}^3\left[{\overline q_{i}}^{A} {(\sigma^I)_A}^B
q_{iB}\right]^2
+ |\phi^{(ab)} q_{iA}|^2  \label{effpot2}
\end{equation}

The fermion-boson coupling can be written as
\begin{equation}
 \bar{\psi}_{ia} \phi^{(ab)} \psi_{ib}
+ {q_{iA}} \chi^{aA} \psi_{ia} + {\overline q_{iA}} \overline \chi^{aA}
\overline \psi_{ia}
\end{equation}

Finally, the full Lagrangian in manifestly $SU(2)_L\times SU(2)_R\times
SU(N_f)$ invariant notation is
\begin{eqnarray}
 \mathcal{L}_S = &&\frac{1}{g^2} \left[ {1\over 4}F_{\mu\nu}^2 + |\partial_\mu \phi^{(ab)}|^2
+ \chi^{aA} \sigma^\mu\partial_\mu \chi_{aA} \right]
 \nonumber \\ && + |\nabla_\mu q_{iA}|^2 + {1\over 2}\sum_I
\left[{\overline q_{i}}^{A} {{(\sigma^I)}_A}^B q_{iB}\right]^2
+ |\phi^{(ab)} q_{iA}|^2 \nonumber \\ &&+{\overline \psi_i}^{a}\sigma^\mu\nabla_\mu
\psi_{ia}+ {\overline \psi_{ia}} \phi^{(ab)} \psi_{ib}
+ {q_{iA}} \chi^{aA} \psi_{ia} + {\overline q_{iA}} \overline \chi^{aA}
\overline \psi_{ia}. \label{full}
\end{eqnarray}
The complete field content, along with their transfromations under $SU(2)_L\times SU(2)_R\times SU(N_f)$, is
\begin{itemize}
\item the gauge field $A_\mu$, transforming as (${\bf 1}$, ${\bf 1}$, 1) (as we will see in Section~\ref{sec:coulomb}
it is convenient to represent this by a dual scalar $\Sigma$, and the monopole operator $\hat{q} \sim e^{2 \pi i \Sigma}$),
\item the matter complex scalars $q_{iA}$,
transforming as
$({\bf 1},{\bf 2}, N_f)$,
\item the matter two-component Dirac fermions $\psi_{ia}$, transforming as $({\bf
2},{\bf 1}, N_f)$,
\item the gauginos (also two-component Dirac fermions)
$\chi_{aA}$, transforming as $({\bf 2},{\bf 2}, 1)$, and
\item the real scalars $\phi^{(ab)}$,
transforming as $({\bf 3},{\bf 1}, 1)$.
\end{itemize}
The first 3 fields above, are direct analogs of the fields present in theories of deconfined critical points
in doped antiferromagnets as described by Eq.~(\ref{fhpsi}).

\section{Supersymmetric phase diagram}

The exact phase diagram (moduli space) for the $\mathcal{N}=4$ theory is obtained by
minimizing the effective potential in Eq.~(\ref{effpot}) or (\ref{effpot2}) --- this potential is protected
by supersymmetry. Note that the minimum energy is always $V=0$, and this equation then specifies the
structure of the moduli space. The low energy excitations above the vacuum are described by a ``sigma model''
on this moduli space, and the gradient terms of this sigma model specify the ``metric'' on moduli space.
Because of the special nature of the effective potential
(\ref{effpot2}), the moduli space is not simply determined by the structure of the broken symmetry (as it is
in non-supersymmetric sigma models). Rather, the target space (moduli space) is a measure of the space
along which the effective potential remains flat, and this has additional constraints imposed by supersymmetry.

For $N_f > 1$, the solution of $V=0$ defines a space with two distinct branches, which will be the Coulomb
and Higgs branches. These meet at a singular point (the ``origin''), which specifies a CFT separating these branches.
For $N_f =1$, there is only branch, which will be identified as the Coulomb branch, and so no phase transition.
Nevertheless, as we will see below in Section~\ref{sec:coulomb}, there is a dual representation of the theory on this branch which
matches closely with the Dasgupta-Halperin duality of the Abelian Higgs model, as already indicated in Eq.~(\ref{zzss}).

The ``Coulomb'' branch of the theory has $\phi^{(ab)} \neq 0$ and so we must have
$q_i=\tilde{q}^i = 0$ to reach $V=0$. This branch breaks the dual symmetry $\widetilde{U}(1)$, and also
$SU(2)_L$, while $SU(N_f)$ and $SU(2)_R$ are preserved. It is easily seen that the matter fields
are massive in fluctuations about any point on this branch. As will be explained in Section~\ref{sec:coulomb},
classically, the Coulomb branch moduli space is ${\mathbb R}^3\times S^1$, parameterized
by $\phi^{(ab)}$ together with the dual photon $\Sigma$. We will explore the structure of the
quantum fluctuation corrections to the moduli space in Section~\ref{sec:coulomb}:
the Coulomb branch
moduli space is deformed to a Taub-NUT space \cite{nut} of NUT charge $N_f$,
in which the circle is nontrivially fibered over the ${\mathbb R}^3$.
On any sphere surrounding the origin of the ${\mathbb R}^3$, the circle bundle over the sphere has degree $N_f$,
and its total space is $S^3/{\mathbb Z}_{N_f}$. Near the origin, the quantum corrected Coulomb
branch moduli space looks like ${\mathbb R}^4/{\mathbb Z}_{N_f}$ rather than ${\mathbb R}^3\times S^1$.

The ``Higgs'' branch
has $q_{iA} \neq 0$ and $\phi^{ab} = 0$, and is present only for $N_f > 1$.
Now the $SU(N_f)$ and $SU(2)_R$ symmetries are broken, while $\widetilde{U} (1)$ and $SU(2)_R$ are preserved.
Comparing with the broken symmetries in the Coulomb branch, we observe that the broken and preserved symmetries are
exactly interchanged, and so a direct transition between them is a non-LGW transition. The moduli space of the Higgs
branch will be described in Section~\ref{sec:higgs}; in this case a classical ({\em i.e.\/} tree-level) analysis
of $\mathcal{L}_S$ yields the correct structure, and quantum corrections are not as important as in the Coulomb
branch.

The two branches meet at the CFT at the origin of the moduli space $\phi^{(ab)}=0$ and $q_{iA} = 0$.
For $N_f = 2$, we will see in Sections~\ref{sec:coulomb} and~\ref{sec:higgs} that the moduli metrics of the Coulomb
and Higgs branches are identical to each other near the origin, demonstrating the self-duality of this case.

\section{The Coulomb branch}
\label{sec:coulomb}

Because $\phi^{(ab)} \neq 0$, the matter fields $q_{iA}$ and $\psi_{ia}$
are massive. So let us integrate these fields out examine the structure of the effective
action at low momenta. The discussion below is an elaborated version of arguments
by Seiberg and Witten \cite{sw}.

We first examine the terms induced in the effective action for the gapless bosonic modes
on moduli space: these are the real scalars $\phi^{(ab)}$, and the gauge field $A_\mu$.
It is convenient to write the scalars in vector notation as $\phi^\alpha \tau^{(ab)}_\alpha$,
where $\tau_\alpha$ are the Pauli matrices with $\alpha = x,y,z$.
Then the induced effective action for the scalars and the gauge field $A_\mu$ is
\begin{equation}
\delta S_{\phi,A} = N_f \mbox{Tr}_A \ln \left[ - \nabla_\mu^2 + \phi^{\alpha 2} \right] - N_f
\mbox{Tr}_a \ln \left[  \sigma^\mu \nabla_\mu + \phi^{\alpha} \tau_\alpha \right]
\end{equation}
For constant $\phi^\alpha$, independent of spacetime co-ordinate $x$, it is now evident that $\delta S_{\phi,A} = 0$,
because the spectra of the two operators co-incide. Therefore, there is no renormalization of the $\phi^\alpha$ superpotential.

To allow for $x$ dependence, we write
\begin{equation}
\phi^\alpha (x) = |\phi| n^\alpha + \delta\phi^\alpha (x)
\end{equation}
where $|\phi|$ is a constant, $n^\alpha$ is a constant unit vector, and $\delta\phi^\alpha (x)$ is a spacetime
dependent fluctuation. Now we can expand in powers of $\delta\phi^\alpha$ and $A_\mu$.
The lowest order terms
are of the form
\begin{equation}
\delta S_{\phi,A} = N_f \int \frac{d^3 p}{8 \pi^3} \left[
\frac{1}{2} K_{\alpha\beta} (p) \delta \phi^\alpha (-p) \delta\phi^\beta (p) + \frac{1}{2} G_{\mu\nu} (p)  A_\mu (-p) A_\nu (p) + \ldots \right]
\end{equation}
The linear coupling between $\delta\phi_\alpha$ and $A_\mu$ is easily shown to vanish.
We now list the expressions for the kernels, initially writing down the contributions of the bosonic and fermionic loops separately
\begin{eqnarray}
K_{\alpha\beta} (p) &=& \int \frac{d^3 q}{8 \pi^3} \left[\frac{ - 8 n_\alpha n_\beta |\phi|^2}{(q^2 + |\phi|^2)(
(q+p)^2 + |\phi|^2)} + \frac{4\delta_{\alpha\beta}}{(q^2 + |\phi|^2)} \right] \nonumber \\
&~&~~~~~~~~~~~~~+ \int \frac{d^3 q}{8 \pi^3} \left[\frac{ - 4 q.(q+p) \delta_{\alpha\beta} + 4 |\phi|^2 (2 n_\alpha n_\beta - \delta_{\alpha\beta})}{(q^2 + |\phi|^2)(
(q+p)^2 + |\phi|^2)}  \right] \nonumber \\
&=& 2 p^2 \delta_{\alpha\beta} \int \frac{d^3 q}{8 \pi^3} \frac{1}{((q+p/2)^2 + |\phi|^2)((q-p/2)^2 + |\phi|^2)} \nonumber \\
&=& \delta_{\alpha\beta} \frac{p^2}{4 \pi |\phi|} ~~~~~\mbox{for $|p| \ll |\phi|$}
\end{eqnarray}
Note that $K_{\alpha\beta} (0)=0$, as expected from the vanishing of the renormalization of the superpotential.
For the kernel of the gauge field, we obtain
\begin{eqnarray}
G_{\mu\nu} (p) &=& \int \frac{d^3 q}{8 \pi^3} \left[\frac{ - 2(2 q_\mu + p_\mu)(2q_\nu + p_\nu)}{(q^2 + |\phi|^2)(
(q+p)^2 + |\phi|^2)} + \frac{4\delta_{\mu\nu}}{(q^2 + |\phi|^2)} \right] \nonumber \\
&~&~~~~~~~~~~~~~+ 4\int \frac{d^3 q}{8 \pi^3} \left[\frac{ 2q_\mu q_\nu + p_\mu q_\nu + p_\nu q_\mu -\delta_{\mu\nu} (q.(q+p) + |\phi|^2)
}{(q^2 + |\phi|^2)(
(q+p)^2 + |\phi|^2)}  \right] \nonumber \\
&=& \frac{(p^2 \delta_{\mu\nu} - p_\mu p_\nu )}{12 \pi |\phi|}+ \frac{(p^2 \delta_{\mu\nu} - p_\mu p_\nu )}{6 \pi |\phi|} ~~~~~\mbox{for $|p| \ll |\phi|$} \nonumber \\
&=& \frac{(p^2 \delta_{\mu\nu} - p_\mu p_\nu )}{4 \pi |\phi|}~~~~~\mbox{for $|p| \ll |\phi|$}
\end{eqnarray}
At higher order, there is a triangle diagram which leads to a cubic term between the $A_\mu$ and
$\delta \phi^\alpha$. This arises only from the fermion loop and leads to a contribution of the form
\begin{eqnarray}
&& A_\mu (-p_1 -p_2) \delta \phi^\alpha (p_1) \delta \phi^\beta (p_2) \times
\left[ 4 i \epsilon_{\mu\nu\lambda} p_{1 \nu} p_{2 \lambda} \epsilon_{\alpha\beta\gamma} n^\gamma |\phi|
\int \frac{d^3 q}{8 \pi^3} \frac{1}{(q^2 + |\phi|^2)^3} \right] \nonumber \\
&&~~~= \frac{i}{8 \pi |\phi|^2} \epsilon_{\mu\nu\lambda} p_{1 \nu} p_{2 \lambda} \epsilon_{\alpha\beta\gamma} n^\gamma A_\mu (-p_1 -p_2) \delta \phi^\alpha (p_1) \delta \phi^\beta (p_2)
\end{eqnarray}

Putting all these terms together, we have the effective action in the Coulomb branch
\begin{equation}
\begin{aligned}
S_{\phi,A} &= \int d^3 x \left[ \frac{1}{2\widetilde{g}^2} \left[ (\partial_\mu \phi^\alpha)^2
+ (\epsilon_{\mu\nu\lambda} \partial_\nu A_\lambda)^2 \right] + \frac{i N_f}{8 \pi |\phi|^3}
\epsilon_{\mu\nu\lambda} \epsilon_{\alpha\beta\gamma} A_\mu \phi^\alpha \partial_\nu \phi^\beta \partial_\lambda \phi^\gamma \right] \label{sp1}
\\
&= \int d^3x \left[ {1\over 2\tilde g^2} ((\partial_\mu\phi^\alpha)^2 + (\widetilde{F}^\mu)^2)
+{iN_f\over 4\pi} \widetilde{F}^\mu \partial_\mu\phi^\alpha {\cal A}_\alpha\right]
\end{aligned}
\end{equation}
where the renormalized coupling is
\begin{equation}
\frac{1}{\widetilde{g}^2 (x)} = \frac{1}{g^2} + \frac{N_f}{4 \pi \sqrt{\phi^\alpha(x) \phi^\alpha (x)}}
\end{equation}
and $\widetilde{F}^\mu \equiv \epsilon^{\mu\nu\rho} \partial_\nu A_\rho$.
Here, and henceforth, we will consider the coupling $\widetilde{g}$ and $|\phi|=\sqrt{\phi^\alpha(x) \phi^\alpha (x)}$ to be arbitrary functions
of the spacetime co-ordinate $x$.
We have also introduced the Dirac monopole function $\mathcal{A}_\alpha$ on $\phi^\alpha$ space which obeys
\begin{equation}
\frac{\partial \mathcal{A}_\alpha}{\partial \phi^\beta} - \frac{\partial \mathcal{A}_\beta}{\partial \phi^\alpha} = \frac{\epsilon_{\alpha\beta\gamma} \phi^\gamma}{|\phi|^3} \label{dm}
\end{equation}
Then we can verify that
\begin{equation}
\epsilon_{\mu\nu\lambda} \partial_\nu \left[ \mathcal{A}_\alpha \partial_\lambda \phi^\alpha \right]
= \frac{1}{2|\phi|^3} \epsilon_{\mu\nu\lambda} \epsilon_{\alpha\beta\gamma} \phi^\alpha \partial_\nu \phi^\beta \partial_\lambda \phi^\gamma
\end{equation}

Eq.~(\ref{sp1}) defines the bosonic sector of the sigma model on the Coulomb branch of moduli space, expressed
in terms of the real scalar $\phi^\alpha$ and the gauge field $A_\mu$. We can also use similar techniques to obtain the fermionic sector, which would be an effective
action for $\chi_{aA}$. Rather than working this out from the Feynman graph expansion, we choose to determined
the fermion Lagrangian from the bosonic sector by supersymmetry.
It is of the form
\begin{equation}
S_F = \int d^3x \left\{ {1\over 2\tilde g^2}\chi \sigma^\mu \partial_\mu \chi
+{1\over 2}\chi \sigma^\mu \tau^\alpha {\cal V}_{\mu\alpha}(\phi,\widetilde{F}) \chi
+ {\cal O}(\chi^4) \right\}
\end{equation}
where the contraction of
spinor and $SU(2)\times SU(2)$ indices are understood.
The supersymmetry transformations are
\begin{equation}
\begin{aligned}
&\delta \phi^\alpha = \varepsilon^{aA}{(\tau^\alpha)_a}^b \chi_{bA} = \varepsilon \tau^\alpha\chi,\\
&\delta A_\mu = i\varepsilon^{aA} \sigma_\mu \chi_{aA} = i\varepsilon \sigma_\mu\chi, \\
&\delta \chi_{aA} =  \left[ -{(\tau_\alpha)_a}^b \sigma^\mu\partial_\mu \phi^\alpha
+ {\delta_a}^b \sigma^\mu \widetilde{F}_\mu \right] \varepsilon_{bA} = \left[(-\tau_\alpha\sigma^\mu \partial_\mu \phi^\alpha
+\sigma^\mu \widetilde{F}_\mu ) \varepsilon\right]_{aA}.
\end{aligned}
\end{equation}
The supersymmetry variation of the bosonic part of the action is
\begin{equation}
\begin{aligned}
\delta S_{\phi,A} &= \int d^3x \left\{\left( {1\over\tilde g^2} \partial^\mu\phi_\alpha + {iN_f\over 4\pi} {\cal A}_\alpha
\widetilde{F}^\mu \right) \varepsilon \tau^\alpha \partial_\mu\chi\right. \\
&\left.
+\left[ -{\partial_\alpha\tilde g\over \tilde g^3}((\partial\phi)^2+\widetilde{F}^2) + {iN_f\over 4\pi} \partial_\alpha {\cal A}_\beta
\widetilde{F}^\mu\partial_\mu\phi^\beta\right] \varepsilon \tau^\alpha \chi\right.\\
&\left.-i\epsilon^{\mu\nu\rho} \left({1\over \tilde g^2} \widetilde{F}_\rho + {iN_f\over 4\pi} {\cal A}_\alpha
\partial_\rho \phi^\alpha \right) \varepsilon \sigma_\mu \partial_\nu \chi
\right\}\\
&= -\int d^3x \left\{\chi\tau^\alpha \varepsilon \left[-\partial_\mu\left( {1\over\tilde g^2} \partial^\mu\phi_\alpha + {iN_f\over 4\pi} {\cal A}_\alpha
\widetilde{F}^\mu \right) \right.\right. \\
&\left.
\left. -{\partial_\alpha\tilde g\over \tilde g^3}((\partial\phi)^2+\widetilde{F}^2) + {iN_f\over 4\pi} \partial_\alpha {\cal A}_\beta
\widetilde{F}^\mu\partial_\mu\phi^\beta\right] \right.\\
&\left.+ \chi\sigma_\mu \varepsilon \,i\epsilon^{\mu\nu\rho} \partial_\nu\left({1\over \tilde g^2} \widetilde{F}_\rho + {iN_f\over 4\pi} {\cal A}_\alpha
\partial_\rho \phi^\alpha \right)
\right\}
\end{aligned}
\end{equation}
The supersymmetry variation of the fermion action is
\begin{equation}
\begin{aligned}
\delta S_{F} &= \int d^3x \left\{ {1\over 2\tilde g^2} \chi \sigma^\mu \partial_\mu
(-\tau_\alpha \sigma^\nu \partial_\nu\phi^\alpha+\sigma^\nu \widetilde{F}_\nu)\varepsilon
+{1\over 2}\chi \sigma^\mu \partial_\mu\left[{1\over \tilde g^2}
(-\tau_\alpha \sigma^\nu \partial_\nu\phi^\alpha+\sigma^\nu \widetilde{F}_\nu)\right]\varepsilon
\right.\\
&\left. + {\cal V}_{\mu\alpha}(\phi,\widetilde{F})\chi \sigma^\mu \tau^\alpha
(-\tau_\beta \sigma^\nu \partial_\nu\phi^\beta+\sigma^\nu \widetilde{F}_\nu)\varepsilon + {\cal O}(\chi^3) \right\}
\end{aligned}
\end{equation}
By the canceling the term proportional to $\chi\varepsilon$ and $\chi\sigma_\mu\tau_\alpha\varepsilon$, we find
\begin{equation}
\begin{aligned}
{\cal V}_{\mu\alpha} &= {1\over 2}\partial_\alpha ({1\over \tilde g^2})\widetilde{F}_\mu -{i\over 2} \epsilon_{\alpha\beta\gamma}
\partial_\beta({1\over \tilde g^2}) \partial_\mu\phi^\gamma \\
&= {N_f\over 8\pi |\phi|^3}\left(-\phi^\alpha \widetilde{F}_\mu +i\epsilon_{\alpha\beta\gamma}\phi^\beta \partial_\mu\phi^\gamma \right)
\end{aligned}
\end{equation}
One can check that the terms proportional to $\chi\sigma_\mu\varepsilon$ and $\chi\tau_\alpha\varepsilon$
cancel the variation of the bosonic part of the action.

In the $g\to \infty$ limit, we can write the fermion action as
\begin{equation}
S_{F,g\to\infty} = {N_f\over 8\pi} \int d^3x \left[{1\over |\phi|} \chi \sigma^\mu \partial_\mu \chi
+ {1\over 2|\phi|^3} \chi (-\phi^\alpha\tau_\alpha \sigma^\mu \widetilde{F}_\mu + \phi^\alpha \tau_\alpha
\sigma^\mu \partial_\mu \phi^\beta\tau_\beta) \chi + {\cal O}(\chi^4) \right] \label{spf}
\end{equation}

In the remainder of this section, we will rewrite the sigma model in Eq.~(\ref{sp1}) and (\ref{spf}) in a different set of `vortex' variables,
designed to highlight the duality properties.
Decoupling the gauge field kinetic
energy by a Hubbard-Stratonovich field $\hat{A}_\mu$ in Eq.~(\ref{sp1}), we obtain
\begin{equation}
S_{\phi,A} = \int d^3 x \left[ \frac{1}{2\widetilde{g}^2} (\partial_\mu \phi^\alpha)^2
+ \frac{\widetilde{g}^2}{2} \hat{A}_\mu^2 + i A_\mu \epsilon_{\mu\nu \lambda} \partial_\nu \left( K_\lambda + \frac{N_f}{4 \pi}\mathcal{A}_\alpha \partial_\lambda \phi^\alpha \right) \right] \label{sp2}
\end{equation}
Together with the fermion Lagrangian, and
performing the integral over $A_\mu$, we obtain a constraint equation which is solved by a dual field $\Sigma$
to yield the dual action
\begin{equation}
\begin{aligned}
\widetilde{S} &= \int d^3 x \left[ \frac{1}{2\widetilde{g}^2} (\partial_\mu \phi^\alpha)^2
+ \frac{\widetilde{g}^2}{2} \left( \partial_\mu \Sigma - \frac{N_f}{4 \pi}\mathcal{A}_\alpha \partial_\mu \phi^\alpha
-{iN_f\phi^\alpha\over 16\pi|\phi|^3} \chi\tau_\alpha \sigma_\mu \chi \right)^2
\right.\\
&\left.+ {1\over 2\widetilde{g}^2} \chi \sigma^\mu\partial_\mu\chi
+{iN_f\over 16\pi} \epsilon_{\alpha\beta\gamma}{\phi^\alpha\over |\phi|^3}(\partial_\mu \phi^\beta) \chi \tau^\gamma \sigma^\mu \chi
+{\cal O}(\chi^4) \right] \label{sp3}
\end{aligned}
\end{equation}

This is the new form of our sigma model on the Coulomb branch, now expressed in terms of the
real scalar $\phi^\alpha$, a new scalar $\Sigma$, and the Dirac fermions $\chi$. Thus we have exchanged the photon $A_\mu$
for a dual scalar $\Sigma$, and from the arguments in Section~\ref{sec:intro} we can anticipate
that $e^{2 \pi i \Sigma}$ is the monopole operator. We will see this emerge in the analyses below.

\subsection{Duality for $N_f = 1$}
\label{sec:nf1}

We have now assembled the ingredients to illustrate the origin of the key duality relation
in Eq.~(\ref{zzss}) for the $\mathcal{N}=4$ theory for $N_f = 1$.

The crucial and remarkable point is that in the limit where we can take
$\widetilde{g}^2 = 4 \pi |\phi|$, Eq.~(\ref{sp3}) is actually a free field theory. To see this, it is useful
to introduce spherical polar co-ordinates $(\rho, \theta, \gamma)$ in $\phi^\alpha$ space so that
$\phi^\alpha = \rho (\sin \theta \cos \gamma, \sin \theta \sin \gamma, \cos \theta)$, where $\rho$, $\theta$,
and $\gamma$ are functions of $x$. Also, let us choose
\begin{equation}
\mathcal{A}_\alpha = \frac{\sin \theta}{\rho (1 + \cos \theta)} \left( \sin \gamma, -\cos \gamma, 0 \right)
\label{adirac}
\end{equation}
so that Eq.~(\ref{dm}) is obeyed.
Then, we define, the two component complex field $\hat{q}_a$ by
\begin{equation}
\hat{q}_a =  \sqrt{\frac{\rho}{2 \pi}} e^{2 \pi i \Sigma} \left(
\begin{array}{c}
\cos (\theta/2) \\
\sin (\theta/2) e^{i \gamma}
\end{array}
\right), \label{zz}
\end{equation}
From this parameterization, we see that $\Sigma$ lives on a circle with circumference 1;
This is consistent with Eq.~(\ref{sp3}) where the line integral of $\mathcal{A}_\alpha \partial_\mu \phi^\alpha$
is defined modulo the area of the unit sphere, which is $4 \pi$.
The field $\hat{q}_a$ is clearly the bosonic monopole operator.
An explicit computation shows easily that the bosonic part of the action $\widetilde{S}$ in Eq.~(\ref{sp3})
is  the free field theory $S_{\hat{q}}$ for $N_f = 1$ and $\widetilde{g}^2 = 4 \pi |\phi|$, where
\begin{equation}
S_{\hat{q}} = \int d^3 x |\partial_\mu \hat{q}_a |^2 . \label{free}
\end{equation}
Note that under $SU(2)_L\times SU(2)_R$, $\hat{q}_a$ transforms as a $({\bf 2}, 1)$. The $\hat{q}_a$ also carry
a charge under a global $\widetilde{U}(1)_f$ symmetry which is the dual of the $U(1)$ gauge field.
Under this  $\widetilde{U}(1)_f$ symmetry $\Sigma \rightarrow \Sigma + c$, where $c$ is a constant.

A similar analysis can be carried out for the fermionic fields.
These
consist of $\chi_{aA}$ gauginos which transform under $SU(2)_L\times SU(2)_R$ as a $({\bf 2}, {\bf 2})$, and
obey a reality condition. After integrating out the matter fields,
for $N_f=1$, these fermions should be transformed to the superpartners of the $\hat{q}_a$: {\em i.e.\/} they
should become a complex doublet, $\hat{\psi}_A$ which transforms as a $(1, {\bf 2})$ and carries a $\widetilde{U}(1)_f$ charge,
and has a free Dirac Lagrangian.
The resulting dual description is therefore simply that
of a {\em free} $\mathcal{N}=4$ hypermultiplet, $\hat{\mathcal{Q}}$. This is the content of the
 duality relation in
Eq.~(\ref{zzss}).

In the $g\to\infty$ limit, the ${\cal O}(\chi^4)$ terms must vanish because the target
space of the sigma model is locally flat ${\mathbb R}^4$ (away from the origin). We can then write the fermionic terms
in the action $\widetilde{S}$ in Eq.~(\ref{sp3}) as
\begin{equation}\label{ferms}
\begin{aligned}
\widetilde{S}_{F,g\to\infty} &= {N_f\over 8\pi} \int d^3x \left[{1\over |\phi|} \chi \sigma^\mu \partial_\mu \chi
+ {1\over 2|\phi|^3} \chi \phi^\alpha \tau_\alpha
\sigma^\mu \partial_\mu \phi^\beta\tau_\beta \chi \right.\\
&\left.-{i\over 2|\phi|^2}\chi\phi^\beta\tau_\beta
\left({4\pi\over N_f}\sigma^\mu \partial_\mu\Sigma - {\cal A}_\alpha \sigma^\mu \partial_\mu\phi^\alpha \right)
\chi \right]
\end{aligned}
\end{equation}
In terms of
\begin{equation}
\hat{q} =  \sqrt{\frac{\rho}{2 \pi}} e^{2 \pi i \Sigma/N_f} \left(
\begin{array}{c}
\cos (\theta/2) \\
\sin (\theta/2) e^{i \gamma}
\end{array}
\right),
\end{equation}
we have
\begin{equation}
\begin{aligned}
& \phi^\alpha = 2\pi\hat q^\dagger \tau^\alpha \hat q,~~~~|\phi|=2\pi|\hat q|^2,\\
& \phi^\alpha \tau_\alpha = 2\pi(2\hat q \hat q^\dagger - |\hat q|^2), \\
& i|\phi|\left({4\pi\over N_f} \partial_\mu\Sigma - {\cal A}_\alpha \partial_\mu\phi^\alpha\right) = \hat q^\dagger \partial_\mu\hat q
-\partial_\mu \hat q^\dagger \hat q.
\end{aligned}
\end{equation}
Now we can rewrite (\ref{ferms}) as
\begin{equation}
\begin{aligned}
\widetilde{S}_{F,g\to\infty} &={N_f\over 16\pi^2}\int d^3x {1\over |\hat q|^4}\left[
|\hat q|^2\chi \sigma^\mu \partial_\mu\chi - (\chi\partial_\mu \hat q)\sigma^\mu (\hat q^\dagger \chi)
+(\chi \hat q)\sigma^\mu (\partial_\mu\hat q^\dagger \chi)\right]\\
&={N_f\over 8\pi^2}\int d^3x
\left({\chi \hat q\over |\hat q|^2}\right)\sigma^\mu \partial_\mu \left({\hat q^\dagger \chi\over |\hat q|^2}\right)
\end{aligned}
\end{equation}
So we see that $\chi_{aA}$ can be mapped to an $SU(2)_R$ doublet of complex fermions $\psi_A$,
\begin{equation}
\hat \psi = {\sqrt{N_f\over2} }\, {\hat q^\dagger \chi\over 2\pi|\hat q|^2}
\end{equation}
which are free away from the origin of the moduli space $\hat q=0$ with action
\begin{equation}
S_{\hat{\psi}} = \int d^3 x  ~\bar {\hat\psi} \sigma^\mu \partial_\mu \hat \psi . \label{free2}
\end{equation}
Thus Eqs.~(\ref{free}) and (\ref{free2}) show that when $N_f=1$, the moduli space
is smooth and reduces to flat ${\mathbb R}^4$. This establishes the equivalence of SQED-1
to the theory of a free $\mathcal{N}=4$ hypermultiplet.

\section{Higgs branch}
\label{sec:higgs}

We will now consider the sigma model on the Higgs branch of the moduli space.
Here we simply have to take the low energy limit of the Lagrangian $\mathcal{L}_S$
in Eq.~(\ref{full}) about vacuum point where $\phi^{ab} =0$ but $q_{iA}  \neq 0$.
(recall that there is no such vacuum for $N_f = 1$). The analysis is simpler than on the
Coulomb branch, because here we can simply set $\phi^{ab} = 0$, and need not include
the fluctuation contribution of the massive $\phi^{ab}$ fields. The main analysis
needed is to project $\mathcal{L}_S$ onto the low energy sector defined by $V=0$.

We will explicit carry out such an analysis for $N_f = 2$. Our purpose here is to illustrate
that the resulting low energy sigma model on the Higgs branch is in fact identical to the
dual version of the sigma model obtained on the Coulomb branch in Eq.~(\ref{sp3}). This identity
then illustrates the self-duality of the $N_f=2$ case, noted by Intrilligator and Seiberg \cite{is}.

A point on the Higgs branch for $N_f = 2$ has $q_{1A} \neq 0$ and $q_{2A} \neq 0$. The vanishing of the
superpotential implies that
\begin{equation}
{\overline q_{1}}^{A} {{(\sigma^I)}_A}^B q_{1B} = - {\overline q_{2}}^{A} {{(\sigma^I)}_A}^B q_{2B}
\end{equation}
for $I=1,2,3$. Let us choose a parameterization of the solutions of this equation which if formally
similar to Eq.~(\ref{zz}).
\begin{equation}
q_{1A} =  \sqrt{\frac{\rho}{4 \pi}} e^{ \pi i \Sigma_1} \left(
\begin{array}{c}
\cos (\theta/2) \\
\sin (\theta/2) e^{i \gamma}
\end{array}
\right)~~~,~~~
q_{2A} =  \sqrt{\frac{\rho}{4 \pi}} e^{ \pi i \Sigma_2} \left(
\begin{array}{c}
\sin (\theta/2) e^{-i \gamma}\\
-\cos (\theta/2)
\end{array}
\right),
 \label{qq}
\end{equation}
By inserting this parameterization
into the Lagrangian $\mathcal{L}_S$ in
Eq.~(\ref{full}), we obtain the action for the bosonic sector of the Higgs branch sigma model.
The gauge field $A_\mu$
higgses out the combination $\Sigma_1 + \Sigma_2$. So we can set $\Sigma_1 = - \Sigma_2  \equiv \Sigma$,
and then ignore $A_\mu=0$.
The resulting action for $\Sigma$, $\theta$ and $\gamma$ is then found to be {\em identical\/} to the bosonic sector of
Eq.~(\ref{sp3})
for $N_f =2$, with the parameterization for $\mathcal{A}_\alpha$ in Eq.~(\ref{adirac}). Thus, as claimed above, the
metric on the moduli spaces of the Higgs and Coulomb branches are identical in the limit $g \rightarrow \infty$.
This identity also explains why the symmetry is enhanced from $\widetilde{U}(1)$ to $\widetilde{SU}(2)$
as we approach the singular CFT point on the moduli space.

For general $N_f$, the Higgs branch
moduli space is a $4(N_f-1)$ real dimensional hyperk\"ahler manifold, which is described as a
hyperk\"ahler quotient ${\mathbb C}^{2N_f}////U(1)$. The latter is explicitly given by
\begin{equation}
\begin{aligned}
&\sum_{i=1}^{N_f} |q_i|^2 - \sum_{i=1}^{N_f} |\tilde q^i|^2 = \zeta_{\mathbb R},\\
& \sum_{i=1}^{N_f} q_i \tilde q^i = \zeta_{\mathbb C},
\end{aligned}
\end{equation}
and modded out by the $U(1)$ action on $q_i$ and $\tilde q^i$ with charges $(+1,-1)$. Here
$\vec\zeta=(\zeta_{\mathbb R},
\zeta_{\mathbb C})$ are the ${\cal N}=4$ Fayet-Illiopoulos parameters, which we have previously
set to zero. In particular, when $N_f=2$, the Higgs branch moduli space is known as the
$A_1$ asymptotically locally Euclidean (ALE) space. For $\vec\zeta=0$ it is
${\mathbb C}^2/{\mathbb Z}_2$, and when $\vec\zeta\not=0$ the ${\mathbb Z}_2$ orbifold
singularity is replaced (resolved) by a $\mathbb{CP}^1$. The Higgs branch moduli space does not
receive quantum corrections, and the classical moduli space is exact. This is because the gauge coupling
can be promoted to a vector multiplet, while the Higgs branch is parameterized by hypermultiplets. The two
decouple at the level of kinetic terms in the low energy effective Lagrangian \cite{plesser}.

\section{Conclusions}
This paper has attempted to straddle the boundaries of two fields, by connecting theories of quantum phase
transitions in square lattice antiferromagnets to dualities of supersymmetric field theories in 2+1 dimensions.
We highlighted to the common physical ideas behind duality mappings in these fields, and so found
explicit examples of quantum phase transitions which do not obey the Landau-Ginzburg-Wilson paradiagm.
Our analysis was restricted to theories with $\mathcal{N}=4$ supersymmetry, where the correspondence
was the simplest. Closely related duality mappings are also available with smaller amounts of supersymmetry \cite{aharony},
at the cost of some additional complexity.

We conclude by noting that another recent example of the parallel developments in the theories of two-dimensional antiferromagnets
and dualities in supersymmetric gauge theories can be found in the close similarities of the theories
and duality mappings described in Refs.~\onlinecite{xus} and~\onlinecite{jaff}.

\acknowledgements
S.S. would like to thank S. Hartnoll and M. Strassler for very useful discussions. We also thank F.~Nogueira
for valuable comments on the manuscript.
This research was supported by the NSF under grant DMR-0757145. X.Y. was supported by a Junior Fellowship
from the Harvard Society of Fellows.

\end{document}